\documentclass[10pt,twocolumn,letterpaper]{article}

\usepackage{cvpr}
\usepackage{times}
\usepackage{epsfig}
\usepackage{graphicx}
\usepackage{caption}
\usepackage{subcaption}
\usepackage{xcolor}
\usepackage{amsmath}
\usepackage{amssymb}
\usepackage{booktabs}
\usepackage{dblfloatfix}
\usepackage{enumitem}
\usepackage[pagebackref=true,breaklinks=true,letterpaper=true,bookmarks=false]{hyperref}

\makeatletter
\renewcommand{\paragraph}{%
  \@startsection{paragraph}{4}%
  {\z@}{0.9ex \@plus 1.2ex \@minus 0.3ex}{-1em}%
  {\normalfont\normalsize\bfseries}%
}
\makeatother
\setlist{topsep=2mm, itemsep=0.5mm, parsep=0mm}

\cvprfinalcopy 

\def\cvprPaperID{1898} 

\begin{document}

\title{A Deep Recurrent Framework for Cleaning Motion Capture Data}

\author{
Utkarsh Mall$^{1, 2}$
\and
G. Roshan Lal$^{1, 3}$
\and
Siddhartha Chaudhuri$^{1, 4}$
\and
Parag Chaudhuri$^1$
\and
\\
$^1$IIT Bombay \,\,\,\,\,\, $^2$Cornell University \,\,\,\,\,\, $^3$University of Wisconsin-Madison \,\,\,\,\,\, $^4$Adobe Research
}

\maketitle

\begin{figure*}
  \includegraphics[width=1.0\textwidth]{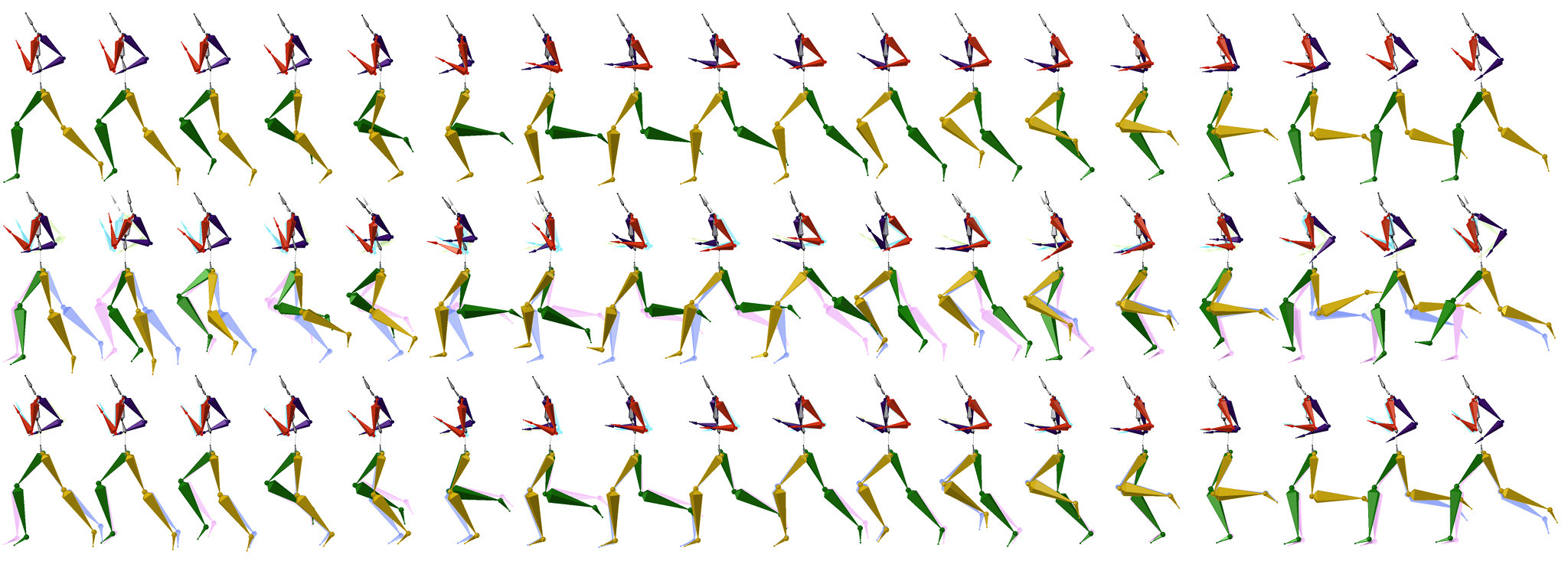}
  \caption{Our deep recurrent motion-cleaning framework takes a base motion (top row) corrupted by substantial noise (middle row) and robustly recovers the underlying signal (bottom row). In the lower rows, a faded image of the base motion is superimposed for reference: note that the cleaned motion (bottom) closely matches the base motion. The method can be trained on a heterogenous mix of actions, handles a variety of noise and gap distributions, does not require manual tuning or specification of the noise model, and operates in a real-time, streaming setting.}
  \label{fig:teaser}
  \vspace{-2mm}
\end{figure*}

\begin{abstract}
\vspace{-2mm}
We present a deep, bidirectional, recurrent framework for cleaning noisy and incomplete motion capture data. It exploits temporal coherence and joint correlations to infer adaptive filters for each joint in each frame. A single model can be trained to denoise a heterogeneous mix of action types, under substantial amounts of noise. A signal that has both noise and gaps is preprocessed with a second bidirectional network that synthesizes missing frames from surrounding context. The approach handles a wide variety of noise types and long gaps, does not rely on knowledge of the noise distribution, and operates in a streaming setting. We validate our approach through extensive evaluations on noise both in joint angles and in joint positions, and show that it improves upon various alternatives.
\vspace{-3mm}

\end{abstract}

\section{Introduction}
\label{sec:introduction}

Data obtained from motion capture (mocap) is used for a variety of purposes, including sports, animation, robotics and medicine. The data is obtained by tracking the movements of (typically) human performers, in either a marker-based~\cite{vicon2017} or a markerless~\cite{organic2017} setup. This data is often noisy and incomplete because of errors present in measurement, tracking or pose reconstruction. This occurs due to several reasons such as calibration error, sensor noise, poor sensor resolution, incorrectly affixed markers or occlusion due to body parts or clothing. Much effort is then spent in cleaning this motion data prior to use. Most often the noisy segments are identified manually, and a particular cleaning technique is applied locally. A few examples of such techniques are interpolation via spline fitting, applying simple smoothing kernels~\cite{vicon2017}, inferring motion of missing joints from neighboring joints on rigid segments~\cite{li2010, aristidou2013}, applying kinematic or geometric constraints~\cite{herda2000, hornung2005}, and applying various filtering techniques~\cite{lou2010, burke2016}. This variety of approaches attests that cleaning motion capture data is a crucial process, and the data is otherwise unusable.

In this paper, we present a deep recurrent framework for cleaning motion capture data (Figure \ref{fig:teaser}). Our method handles both noisy signals and signals with missing data or gaps. After the supervised training phase, it requires no further annotation other than knowledge of where the gaps occurred, corresponding to intervals when markers went missing. The approach is {\em not noise-specific or action-specific}: it can be trained on any type of noise, and a heterogeneous mix of action types. The noise distribution can be completely unknown as long as noisy and clean motion pairs are available for training. Further, the model adapts on-the-fly to the test action type (e.g. ``walk'', ``run'', ``jump'') as long as the training includes some (unlabeled) examples of this type. Finally, the approach can operate in a streaming setting, processing mocap data in real time as it arrives instead of needing to look at the complete motion clip at once.

Such a framework faces several challenges. {\em First}, different joints of a tracked character move at different and varying speeds, especially across different phases/types of actions (e.g. foot swing vs strike, or walk vs jump). Hence, na\"ive smoothing risks either blurring out sharp movements or not removing enough noise. {\em Second}, if the noise is significant, then an isolated short interval of the motion contains few cues to reliably recover the underlying clean signal by model fitting. {\em Third}, noise distributions can be unknown, arbitrary, and have non-zero mean. This makes handcrafting an approach for blind noise removal distinctly non-trivial. {\em Fourth}, gaps created by missing markers can be long enough to omit complex short-duration motions: such gaps cannot be filled by simple interpolation.

We address these challenges with a recurrent, bidirectional, long short-term memory (B-LSTM) neural network architecture \cite{graves2005} that predicts an adaptive denoising filter for each joint in each frame. We call our network architecture {\em Encoder-Bidirectional-Filter (EBF)}. By setting up correlations between joints and exploiting temporal coherence, the EBF network can robustly infer the current phase of motion and its frequency, and decide on an appropriate filter. This avoids over or under-blurring. Since it maintains persistent, hidden state to track the motion over time, the network can reliably contextualize the current frame even in the presence of huge amounts of noise. Given suitable training data in the form of noisy and clean motion pairs, it can be trained to remove a wide variety of noise types from a wide variety of actions without any prior knowledge of the noise distribution. For non-zero mean noise, a debiasing term can also be learnt to remove bias, even in cases where the bias is (deterministically) varying.

For a signal that has both noise and gaps, we preprocess the signal with a second B-LSTM network that synthesizes missing frames from surrounding context. This {\em EBD} network, a bidirectional variant of the Encoder-Recurrent-Decoder (ERD) architecture of Fragkiadaki et al.~\cite{fragkiadaki2015}, is trained with data augmentation and dropout to be robust to long-duration synthesis. It fills a gap with a sequence that captures the correct trend but is still somewhat noisy. (As we show in our experiments, the EBD architecture is not solely sufficient to produce a clean signal.) The complete, noisy motion is then cleaned with our first EBF network.

We show through extensive evaluations that our framework successfully cleans both small and large amplitude noise, and fills gaps, better than a variety of baselines. We also develop a benchmark dataset of a variety of motion types~\cite{cmu2017} corrupted by various types of synthetic noise and gaps. We will put this dataset, as well as code and trained models, in the public domain.

\section{Related Work}
\label{sec:background}

\begin{figure*}[t!]
\centering
  \includegraphics[width=\linewidth]{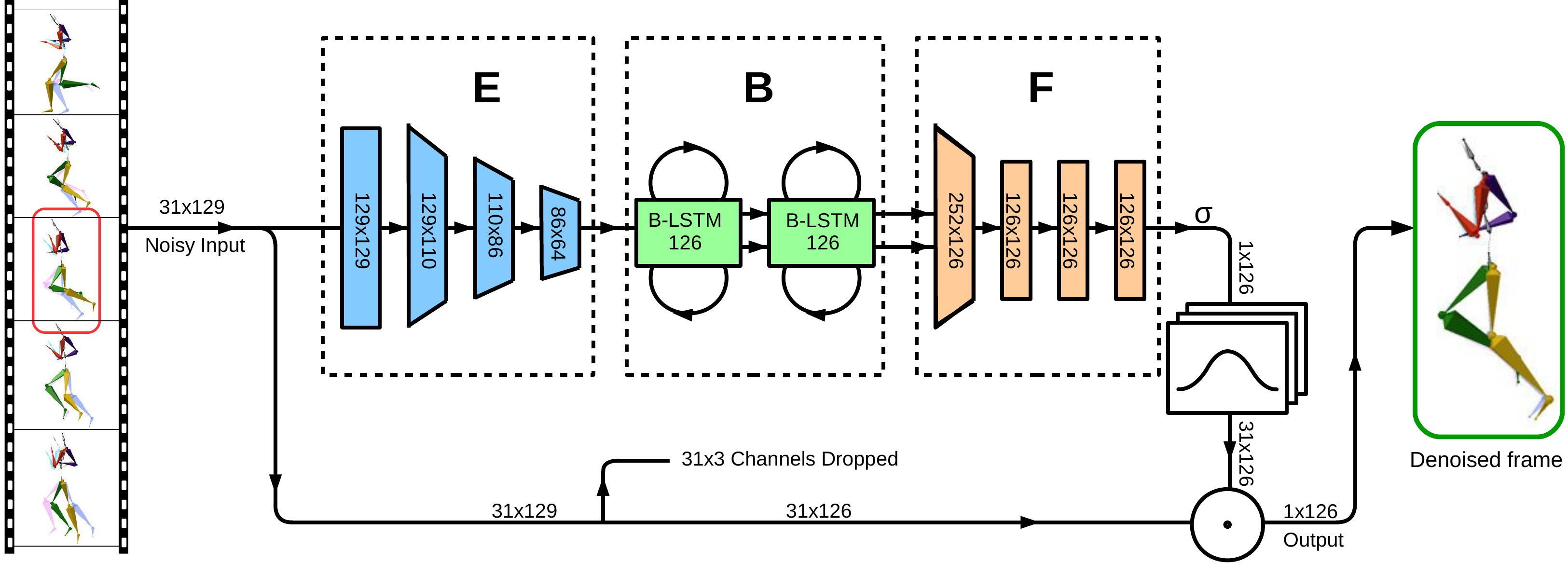}
  \caption{Denoising Pipeline. A mocap frame at time $t$ (circled in red), plus 15 frames each of lookahead and lookback (left), are fed to the network and passed through encoder (E), B-LSTM (B) and filter prediction (F) layers. Each joint channel in the final cleaned frame (right) is a weighted average of corresponding channels in surrounding frames, where the weights are predicted by the network as a Gaussian with adaptive variance. Optional debiasing or fine-grained filter shape prediction components are omitted for clarity. 3 input channels for root angular velocity are used only to contextualize the pose and are not output by the pipeline. Ground truth clean frames are faded and superimposed on the input/output frames for reference.}
  \label{fig:pipeline}
\end{figure*}

Rudimentary motion cleaning techniques that handle both noisy and missing samples exist in software systems available from mocap solution providers like Vicon~\cite{vicon2017} and 3D content creation tools like Maya~\cite{maya2017} or Blender~\cite{blender2017}. These rely on simple interpolation and filtering, along with kinematic constraints, and have to be applied manually to the input motion signals.

In prior research, mocap cleaning has often been combined with marker tracking, labeling and reconstruction. We now discuss a variety of such approaches.

\paragraph{Skeleton based methods.} Herda et al.~\cite{herda2000} and Hornung et al.~\cite{hornung2005} try to improve robustness of marker tracking/labeling in marker-based mocap by using a kinematic skeleton to assist marker reconstruction.
Zordan and Van Der Horst~\cite{zordan2003} also use a fixed skeleton to map markers and resolve joint angle state with a dynamics model.
In these methods, the skeleton-to-marker mapping is fixed and explicitly specified. They can only reconstruct markers that are briefly missing, and are not robust to noise.

\paragraph{Kalman filter based methods.} Kalman filters have been used to track markers by assuming rigid constraints in marker placement~\cite{dorfmuller2003}. Li et al.~\cite{li2010} present the BoLeRO method to enforce bone length constraints in a linear dynamical system. Aristidou and Lasenby~\cite{aristidou2013} use improved tracking but still assume rigid limbs with $3$ markers on each limb. The use of these methods is limited to scenarios where the markers, missing or not, follow these assumptions and the noise model is known a priori.

\paragraph{Dimensionality reduction based methods.} Liu and McMillan~\cite{liu2006} and Park and Hodgins~\cite{park2006} reconstruct missing markers by projecting motion onto its principal components. Burke and Lasenby~\cite{burke2016} combine temporal smoothing with a Kalman filter and low rank matrix completion to learn an effective subspace for filtering. This fixes gaps, but the work is not robust to heavy noise. Akhter et al.~\cite{akhter2012} use a bilinear spatiotemporal basis to factor the motion into shape and trajectory. This does not enforce coherence and hence works only in a narrow linearly approximable regime. Xiao et al.~\cite{xiao2015} use L1-minimization to learn an optimal dictionary for denoising. This method requires solving an expensive optimization problem for each test motion, requires access to the entire motion clip, and is demonstrated only on tiny noise amplitudes ($30$dB SNR, vs our tests with a default of $6$dB). Lou and Chai~\cite{lou2010} learn filter bases from a few clean samples, and then use non-linear optimization to determine filter weights and the clean motion. This does not generalize over a large variety of input motions as separate filter bases need to be learnt for each motion type. It also requires expensive test-time optimization.

\paragraph{Generative models and deep learning based methods.} Taylor et al.~\cite{taylor2007} present a Conditional Restricted Boltzmann Machine that generates motion signals similar to learnt data and can thereby fill gaps. Fragkiadaki et al.~\cite{fragkiadaki2015} present an ERD network to generate human motions that is easier to train and generalizes better than earlier work. It however, produces very short motion sequences of upto $600$ms, and does not address motion cleaning. Du et al.~\cite{du2015} train a hierarchical RNN to recognize actions in motion sequences. Holden et al.~\cite{holden2015, holden2016, holden2017} develop a variety of deep networks to synthesize and edit motion. LSTM variants have been used to denoise speech~\cite{barker2016, mousa2015, weninger2015}.

In contrast to existing literature, our method is not restricted to any particular kind of marker placement, skeleton, noise or motion model, since these are implicitly learnt from data. It is robust to noise and missing samples, and works for a large variety and length of motions, different kinds of unknown large-amplitude noise, and long gaps.

\section{Method}
\label{sec:method}

Our approach leverages a recurrent, bidirectional long short-term memory neural network. We are inspired by the Encoder-Recurrent-Decoder (ERD) architecture of Fragkiadaki et al.~\cite{fragkiadaki2015}, which synthesizes motion frames given a short initial sequence. However, while the ERD architecture can synthesize plausible gross motions, these sequences are not noise-free when conditioned on noisy observations. It captures general motion trends well, but is less adept at removing fine-grained high-frequency noise. Hence, our architecture focuses on outputting a \mbox{\em denoising filter} for the motion in each frame, rather than attempting to directly produce a denoised pose. (Recent work on denoising rendered images has explored filter prediction in a feedforward convolutional framework~\cite{bako2017}.)  As we show in our evaluation, this approach was critical in producing satisfactory results.

Further, the ERD network contains an encoder component to transform the input to a higher-dimensional space to improve performance of the subsequent recurrent layers. In contrast, our network projects the input to a {\em lower}-dimensional manifold, similar to an autoencoder, to constrain processing to a set of plausible configurations. This simplifies the operation of the recurrent network.

Lastly, in contrast to Fragkiadaki et al., whose primary goal was motion synthesis, our recurrent architecture is {\em bidirectional}, allowing us to utilize both past and future context to estimate the best adaptive filters to clean the noisy input. We are able to do this because we have access to the entire (noisy) temporal sequence.

\paragraph{{\bf Architecture.}} Our Encoder-Bidirectional-Filter (EBF) architecture comprises an encoder module (E) to project the input pose (and lookback) to a low-dimensional manifold, a recurrent bidirectional LSTM component (B) to exploit temporal coherence, and a filter prediction module (F) that outputs the smoothing filter to be applied to each joint in the current frame.

The input representation is the vector of $126$ joint angles $\mathbf{x}(t)$ of the skeletal model at time $t$ (measured in frames), plus the $3$ global angular velocities (around the X, Y and Z axes) of the root node of the character, for a total of $129$ parameters. The network consists of $4$ fully-connected layers (Encoder), followed by $2$ bidirectional recurrent layers (B-LSTM), followed by $4$ more fully-connected layers (Filter Prediction). In our default implementation, the network outputs $126$ values, each interpreted as the standard deviation of a Gaussian which will be used to smooth the motion of the corresponding joint in the current frame. We use $\tanh$ as the activation function in the encoder and filter prediction layers, except for the last filter layer, which uses $\exp$ to ensure that the output standard deviations are positive.

The encoder module reduces the input $129$-dimensional pose vector to a compact $64$-dimensional code. This is then processed by the B-LSTM with $15$ frames of lookback and $15$ frames of lookahead (for a frame-rate of $120$ fps). The B-LSTM output is mapped to a $126$-dimensional filter vector by the  network. The overall EBF architecture is illustrated in Figure \ref{fig:pipeline}.

The $3$ extra channels containing the angular velocities for the root joint are assumed to be clean and are used only for disambiguation. To make the training invariant to some simple transformations such as heading changes, we use angular velocities instead of absolute angles in these channels.

For training the network for non-zero mean noise, we modify the existing EBF to predict both Gaussian filter parameters as well as a bias parameter for every channel. This network predicts $126$ standard deviations \mbox{$\mathbf{\sigma}(t) \in \mathbb{R}^{126}$} and $126$ values of bias \mbox{$\mathbf{b}(t) \in \mathbb{R}^{126}$} for each channel at time $t$. The smoothed pose $\overline{\mathbf{x}}(t)$ produced by the network is given by the weighted sum:
\[
  \overline{x_i}(t) \ = \ \left( \sum_{\tau = t - 15}^{t + 15} w_i(t, \tau) x_i(\tau) \right) - b_i(t)
\]
where
\[
w_i(t, \tau) \ = \ \frac{1}{Z} \exp\left( -\frac{(\tau - t)^2}{2 \sigma_i(t)^2} \right)
\]

\begin{figure}[b!]
\centering
  \begin{subfigure}[b]{0.32\linewidth}
    \includegraphics[width=\columnwidth]{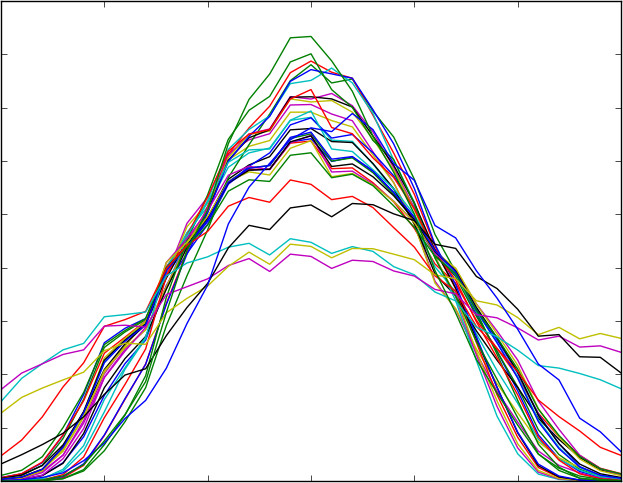}
  \end{subfigure}~~%
  \begin{subfigure}[b]{0.32\linewidth}
    \includegraphics[width=\columnwidth]{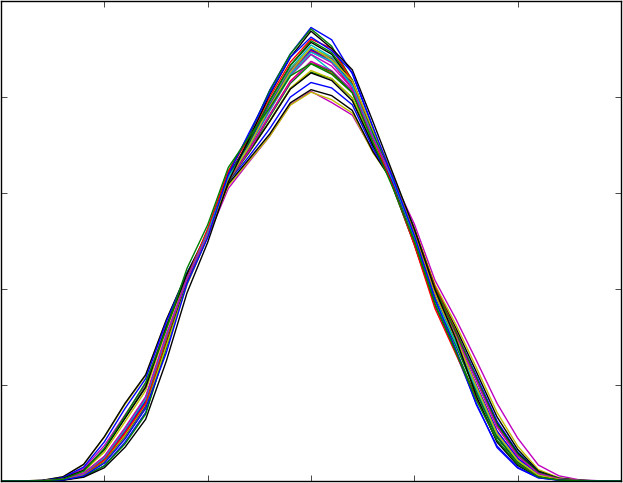}
  \end{subfigure}~~%
  \begin{subfigure}[b]{0.32\linewidth}
    \includegraphics[width=\columnwidth]{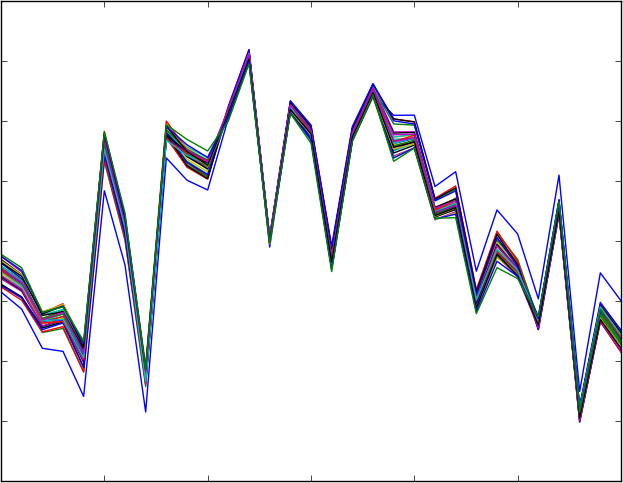}
  \end{subfigure}
  \caption{Filter shapes learnt by an EBF variant that explicitly outputs the distribution of weights (vertical axis) over time (horizontal axis). Each colored curve is the plot of a predicted filter kernel. (Left) Filters for a typical walking motion, covering a range of variances to adapt to the local motion characteristics. (Middle) A tight cluster of constant-variance filters, learnt for a running motion. (Right) The least ``Gaussian-like'' set of filters, learnt for a jump motion. For this experiment, we used the same weight for all joints at a given time, for simplicity. In our main experiments, we directly output a vector of Gaussian filter parameters.}
  \label{fig:expt6}
\end{figure}

\begin{figure*}[t!]
\centering
  \includegraphics[width=\linewidth]{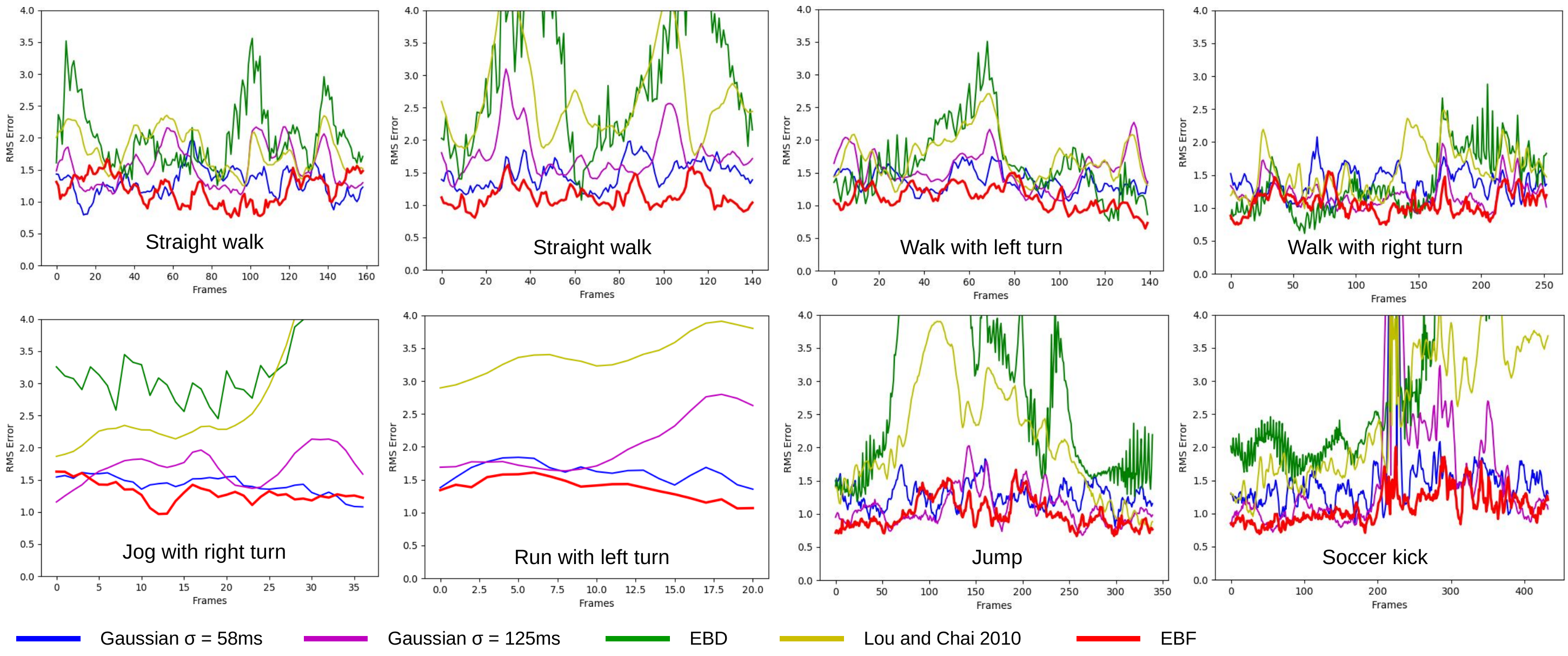}
  \caption{A comparison of five different motion denoising methods on $8$ diverse test motions with 0.5 angular noise from our first holdout set. In each plot, the vertical axis is RMS error over all joint angles, and the horizontal axis is frame number (i.e. time). Our EBF model outperforms Gaussian filters with standard deviations of $58$ and $125$ms respectively, as well as an EBD recurrent network similar to Fragkiadaki et al.~\cite{fragkiadaki2015} and the example-based denoising method of Lou and Chai~\cite{lou2010}. The overall error, averaged over time, for each such motion is shown in Figure \ref{fig:expt1_2}.}
  \label{fig:expt1_1}
\end{figure*}

\begin{figure*}[b!]
\centering
  \begin{subfigure}[b]{0.32\linewidth}
    \includegraphics[width=\columnwidth]{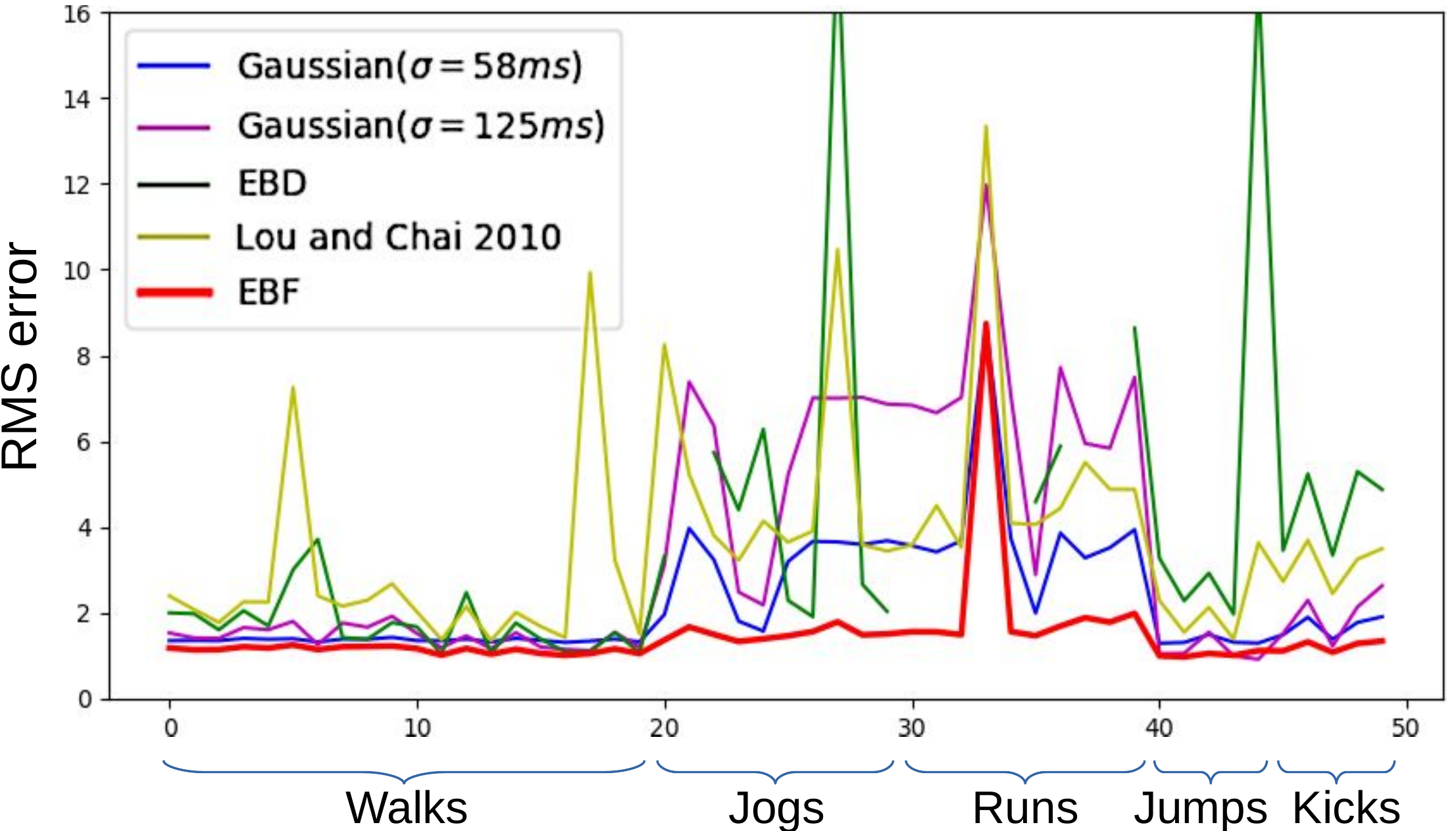}
    \caption{Angular noise ($0.5 \times$ joint std dev)}
    \label{fig:expt1_2_angular}
  \end{subfigure}~%
  \begin{subfigure}[b]{0.32\linewidth}
    \includegraphics[width=\columnwidth]{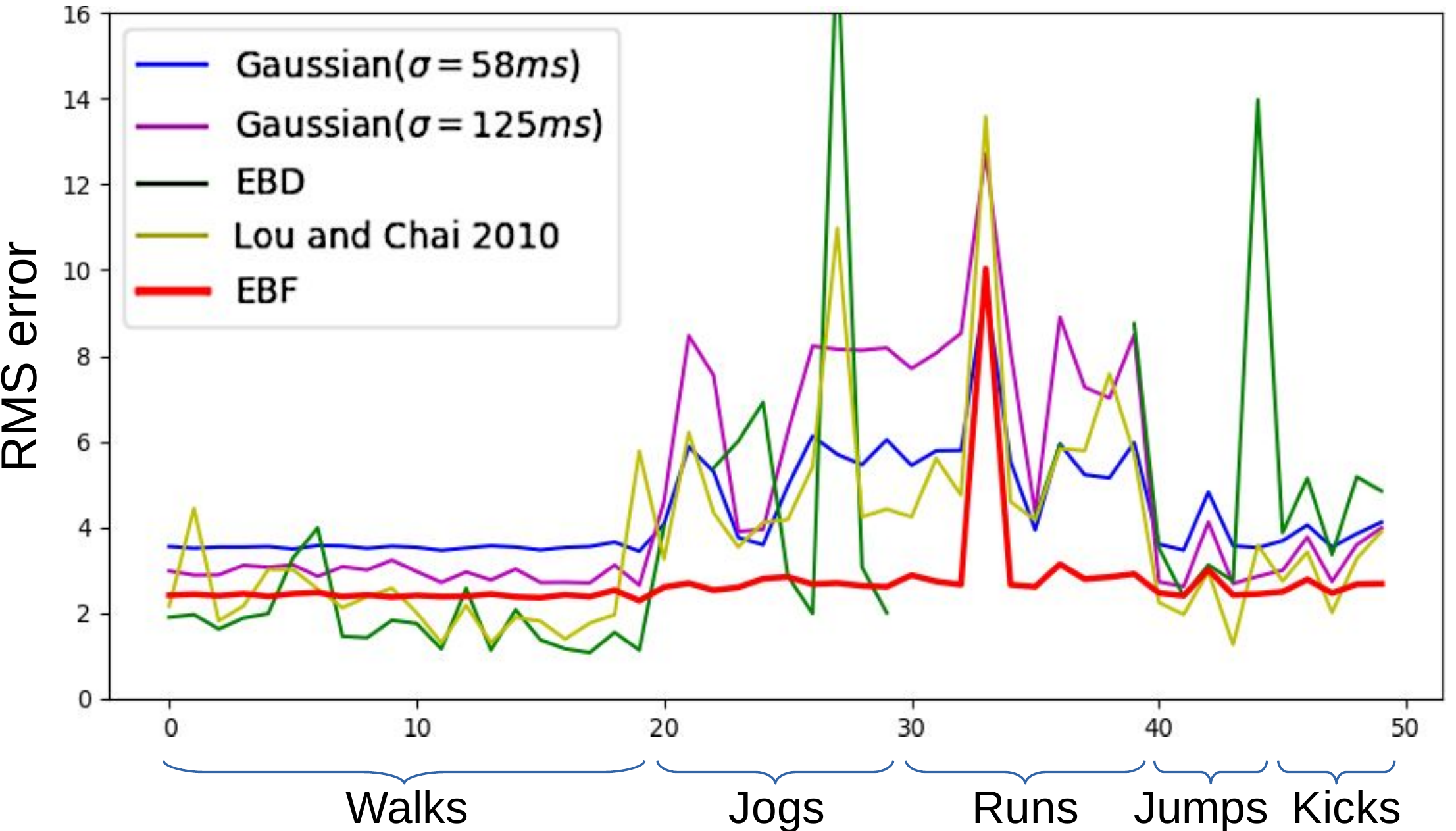}
    \caption{Spatial noise ($0.5$cm)}
    \label{fig:expt1_2_spatial}
  \end{subfigure}~%
  \begin{subfigure}[b]{0.32\linewidth}
    \includegraphics[width=\columnwidth]{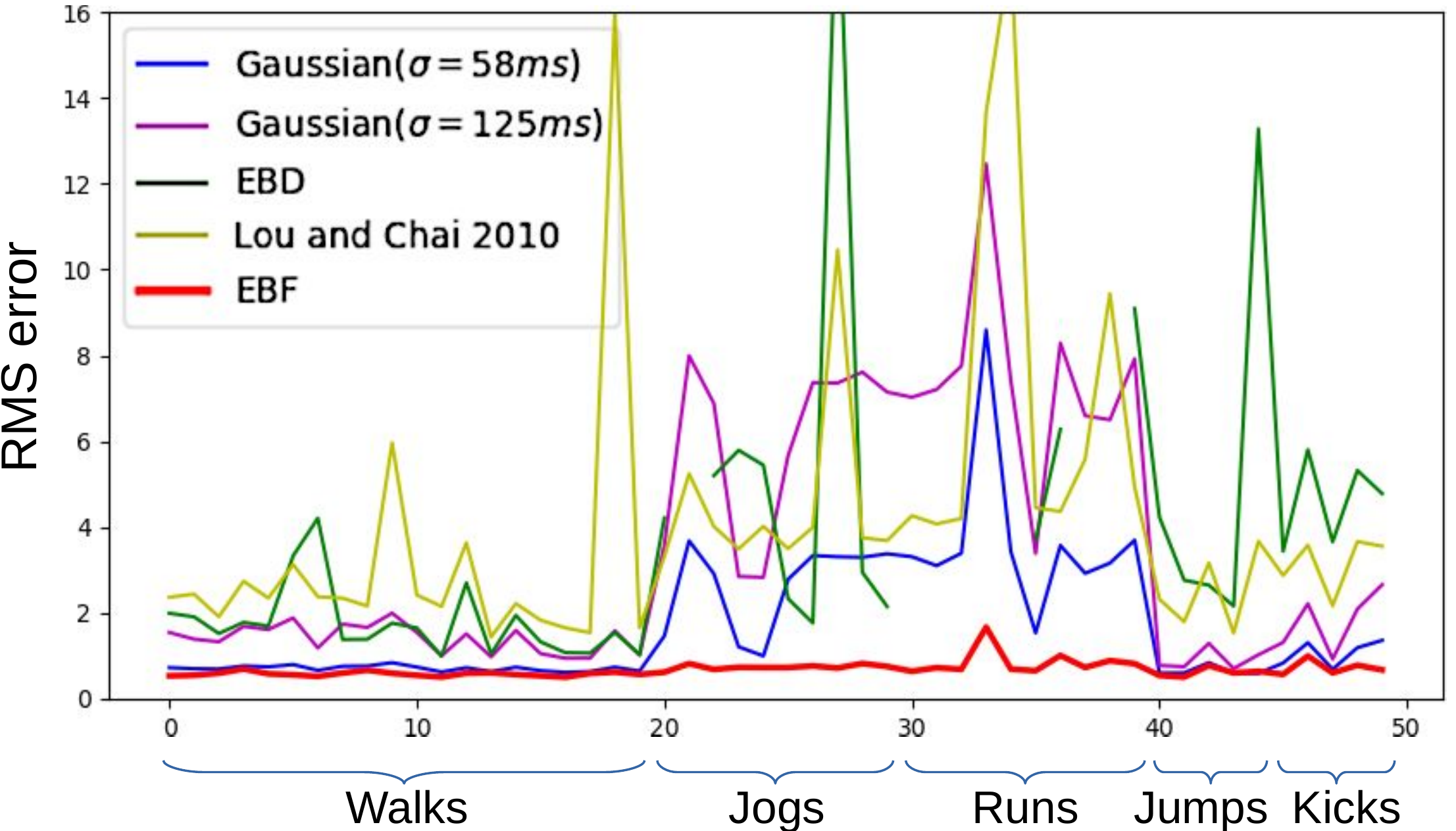}
    \caption{Spatial ($0.5$cm wrist/ankle, $0.1$cm rest)}
    \label{fig:expt1_2_spatial_wristankle}
  \end{subfigure}
  \caption{A comparison of five different motion denoising methods for all the $50$ motions in our test holdouts, numbered $0$ to $49$. For each motion, the RMS error has been averaged over all frames.}
  \label{fig:expt1_2}
\end{figure*}

\noindent Here $Z$ is a normalization term. Our architecture can easily accommodate more fine-grained filter models. For instance, we might directly predict general weight values $w_i(t, \tau)$ over the $31$-frame filtering window. In our experiments with such models, however, we found that the predicted filter shapes were essentially Gaussian (Figure \ref{fig:expt6}). We predicted parameters of Gaussian filters in all our experiments. Nevertheless, the framework can accommodate other filter shapes should the need arise.

\paragraph{{\bf Training.}} The EBF network is trained end-to-end by providing a sequence of noisy frames as input and the corresponding clean ground truth frames as expected output. The network uses L2 loss between motion cleaned with the predicted filters and the ground truth. This setup supports back-propagation since the filtering operation itself is a simple dot product with the $31$-D weight vector for each joint channel (plus a preceding set of exponentiations to compute the weights from the predicted Gaussian standard deviation). We initialize the network weights of the encoder and filter layers by sampling from a uniform distribution over $[-0.5,0.5]$, and the weights of the B-LSTM by sampling from a Gaussian distribution. To regularize, input channels to the B-LSTM are dropped out with a probability of $0.5$. We train the network for $200$ epochs using stochastic gradient descent (SGD) with adaptive moments.

\paragraph{{\bf Gap filling.}} Mocap data can have gaps due to occlusions while tracking markers. For filling in the missing gaps we use an {\em Encoder-Bidirectional-Decoder (EBD)} architecture. The encoder module is identical to that of EBF. The B-LSTM module is nearly identical, except it has fewer neurons in each hidden layer ($32$ vs $126$ in each of the forward and backward directions). The decoder module recovers the $126$-D joint angle vector from the B-LSTM output through $4$ fully-connected layers. The network has a lookback and lookahead of $15$ frames with a stride of $4$ frames (i.e. covering a $60$ frame interval) in each direction. This architecture is directly inspired by the (unidirectional) ERD architecture of Fragkiadaki et al.~\cite{fragkiadaki2015}. It can reconstruct missing channels from the other joint angles by exploiting inter-joint correlations and temporal coherence.

We make the reasonable assumption that the locations of gaps are known (since we can tell when markers went missing). The gap is initially filled with values linearly interpolated from its endpoints, during both training and testing. The EBD learns to replace the linearly interpolated data with the correct sequence of values (non-gap portions at other times or for other joints are left unaltered). The completed sequence is then passed on to the EBF network for denoising.

Like EBF, the EBD network is also trained end-to-end for $200$ epochs using SGD with adaptive moments. The B-LSTM layers have an input dropout of $0.5$. For successful training and robust long range prediction, we found it necessary to augment the training data with many different motions and noise samples, as described in the next section.

\section{Results}
\label{sec:results}

\begin{figure*}[t!]
\centering
  \begin{subfigure}[b]{0.32\linewidth}
    \includegraphics[width=\columnwidth]{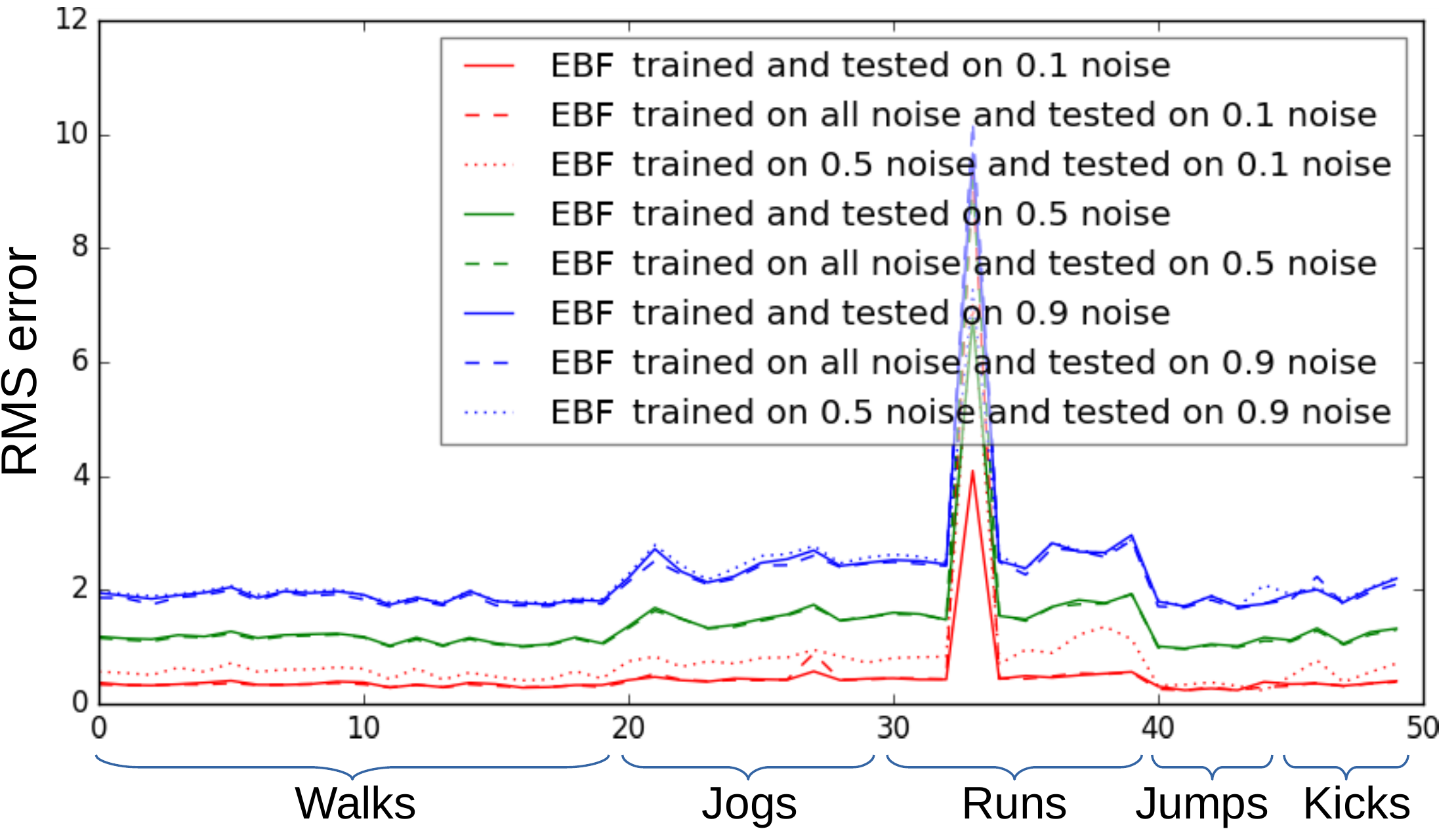}
  \end{subfigure}~%
  \begin{subfigure}[b]{0.32\linewidth}
    \includegraphics[width=\columnwidth]{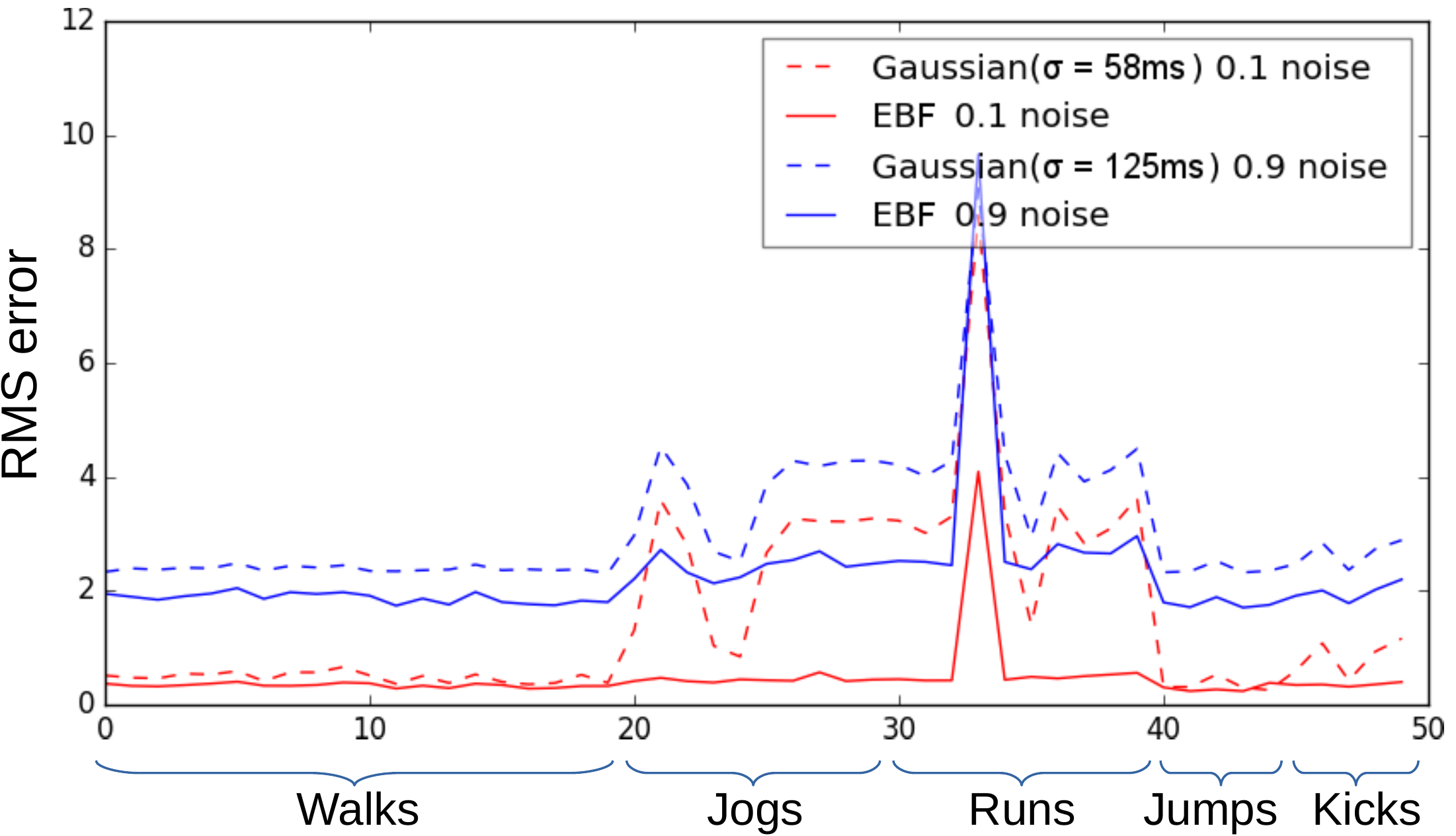}
  \end{subfigure}~%
  \begin{subfigure}[b]{0.32\linewidth}
    \includegraphics[width=\columnwidth]{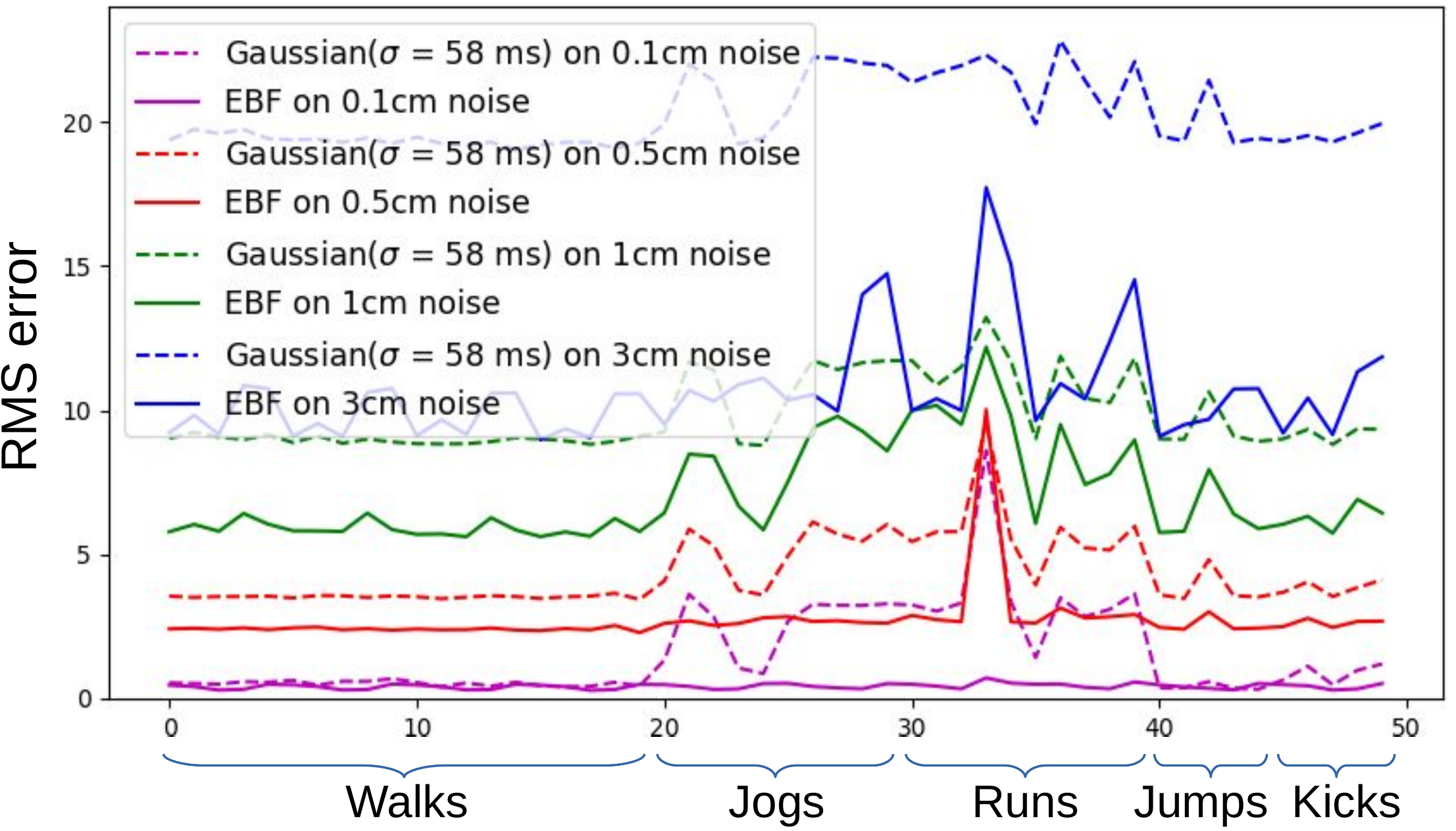}
  \end{subfigure}~%
  \caption{Testing EBF on different noise amplitudes. {\em Left:} The effect of training on all angular noise amplitudes together, vs training on a single one. {\em Middle:} EBF denoising vs Gaussian filtering on $0.1$ and $0.9$ angular noise. {\em Right:} EBF denoising vs Gaussian filtering on $0.1$cm, $0.5$cm, $1$cm and $3$cm spatial noise.}
  \label{fig:expt2}
\end{figure*}

We collected $100$ motions from the CMU mocap database~\cite{cmu2017}, comprising a total of $40$ walks, $20$ jogs, $20$ runs, $10$ jumps and $10$ kicks. Every motion is at $120$ fps and is stored as a BVH file.

\paragraph{Generating noisy train/test data.} We assume that the trajectory of the root of the skeleton is clean. Following standard practice~\cite{lou2010,xiao2015}, we generate noisy variants of motions by applying noise to each joint independently.

The default ``angular'' noise for each joint rotation channel is sampled from a mean-zero Gaussian with standard deviation equal to $0.5$ times the standard deviation of the channel ($6$dB SNR). Hereafter, we refer to this as ``$0.5$ noise''. We also consider motions corrupted by ``spatial'' noise, generated by applying 3D Gaussian noise to joint {\em positions} and optimizing to preserve bone lengths. Spatial noise models motion corruption due to errors in marker location. We test on two types of spatial noise: (a) the same variance at all joints, and (b) larger variance at wrist and ankle end effectors, modeling greater tracking error at fast moving markers. Lastly, we also test with a variety of other synthetic noise types to show the robustness of the method.

We generate missing samples or gaps in the data by sampling from a distribution over gap lengths that decays exponentially with increasing length~\cite{burke2016}. There is a $10\%$ probability of starting a gap in a specific joint channel at a specific frame in the motion. (Note that very few joint channels are likely to have gaps simultaneously under this model -- this is critical for our method since it exploits correlation between joints to reconstruct missing data.)

We use $5$ randomly sampled holdouts of $10$ motions each for cross validation of our framework. In each holdout we ensure there are $4$ walks, $2$ jogs, $2$ runs, $1$ jump and $1$ kick motion, randomly selected from their respective types. For statistical robustness, all our tests, on each type of motion, are repeated $5$ times and the results are averaged across runs. In each run, we vary a motion by adding different noise and gap samples. The noise and gap distributions remain fixed across runs.

\subsection{Denoising Results}

\paragraph{Comparison of denoising methods.} We compare the performance of our EBF model (without bias outputs) for denoising motion capture data with two baseline Gaussian filters, an Encoder-Bidirectional-Decoder (EBD) model similar to Fragkiadaki et al.~\cite{fragkiadaki2015}, and the example-based denoising method of Lou and Chai~\cite{lou2010} (which the authors show to improve upon Kalman and data-driven Kalman filters). First, we plot the variation in RMS error over all joints with time. Figure~\ref{fig:expt1_1} shows these plots for $8$ different test motions. A single EBF model trained over a mix of motion types (red curve) has lower error compared to all other competing methods. Figure~\ref{fig:expt1_2} presents the RMS error averaged over all frames of a motion, for all $50$ motions from the $5$ holdouts. The motions have been grouped by type, for clarity. Again, we see that our EBF model outperforms the other methods.

The average RMS error across all motions in our test set is $2.25$, $3.37$, $3.57$, $3.67$ and {\bf 1.46} for a Gaussian with a standard deviation of $58$ms, a Gaussian with a standard deviation of $125$ms, EBD, Lou and Chai~\cite{lou2010}, and EBF, respectively. The corresponding figures for $0.5$cm spatial noise are  $4.36$, $4.78$, $3.65$, $3.75$ and {\bf 2.72} respectively.

Visualizations of noisy and cleaned motions are shown in the supplementary video.

\begin{figure*}[t!]
\centering
  \begin{subfigure}[b]{0.32\linewidth}
    \includegraphics[width=\columnwidth]{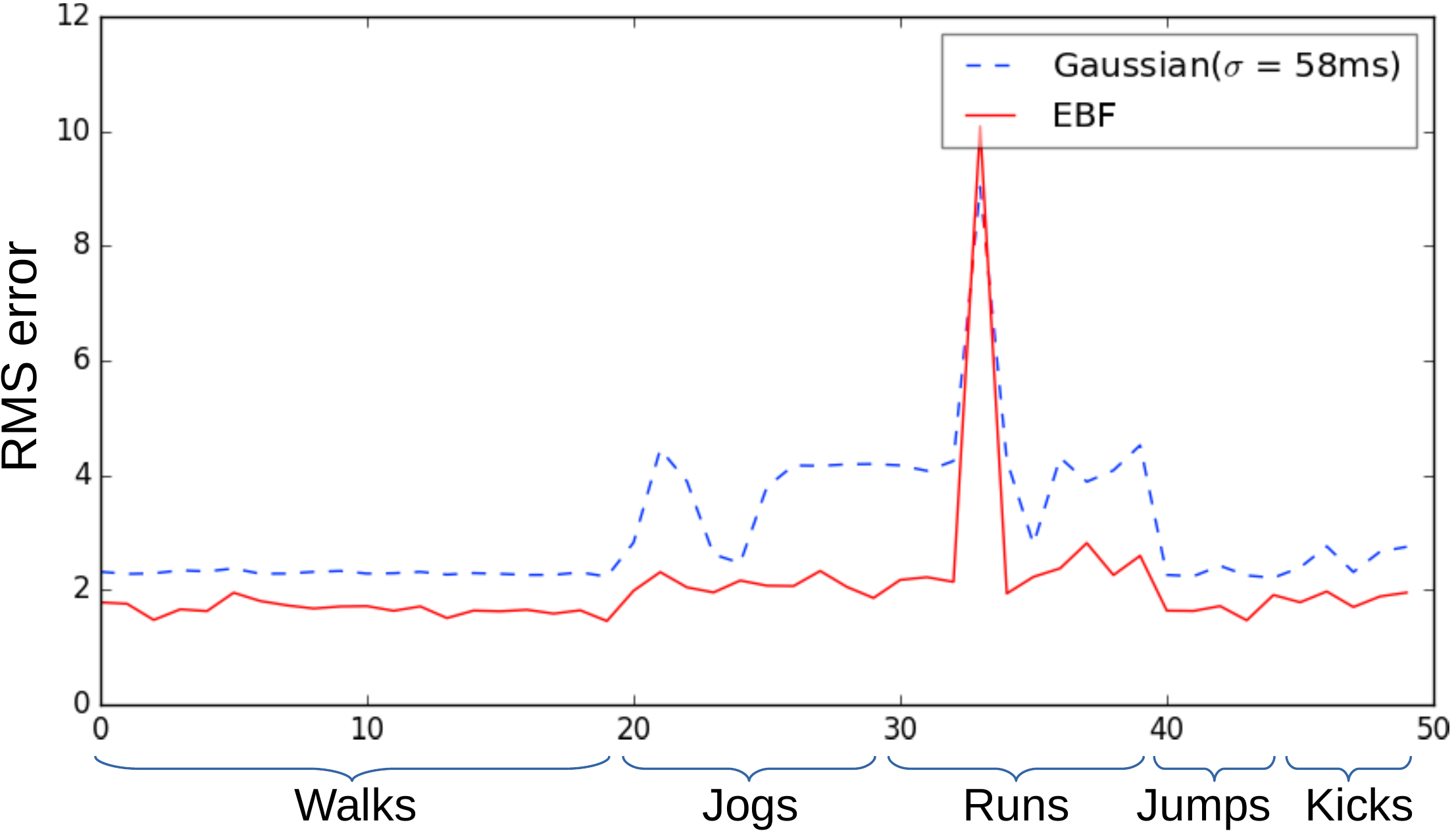}
    \caption{Uniform noise}
    \label{fig:expt3_uniform}
  \end{subfigure}~%
  \begin{subfigure}[b]{0.32\linewidth}
    \includegraphics[width=\columnwidth]{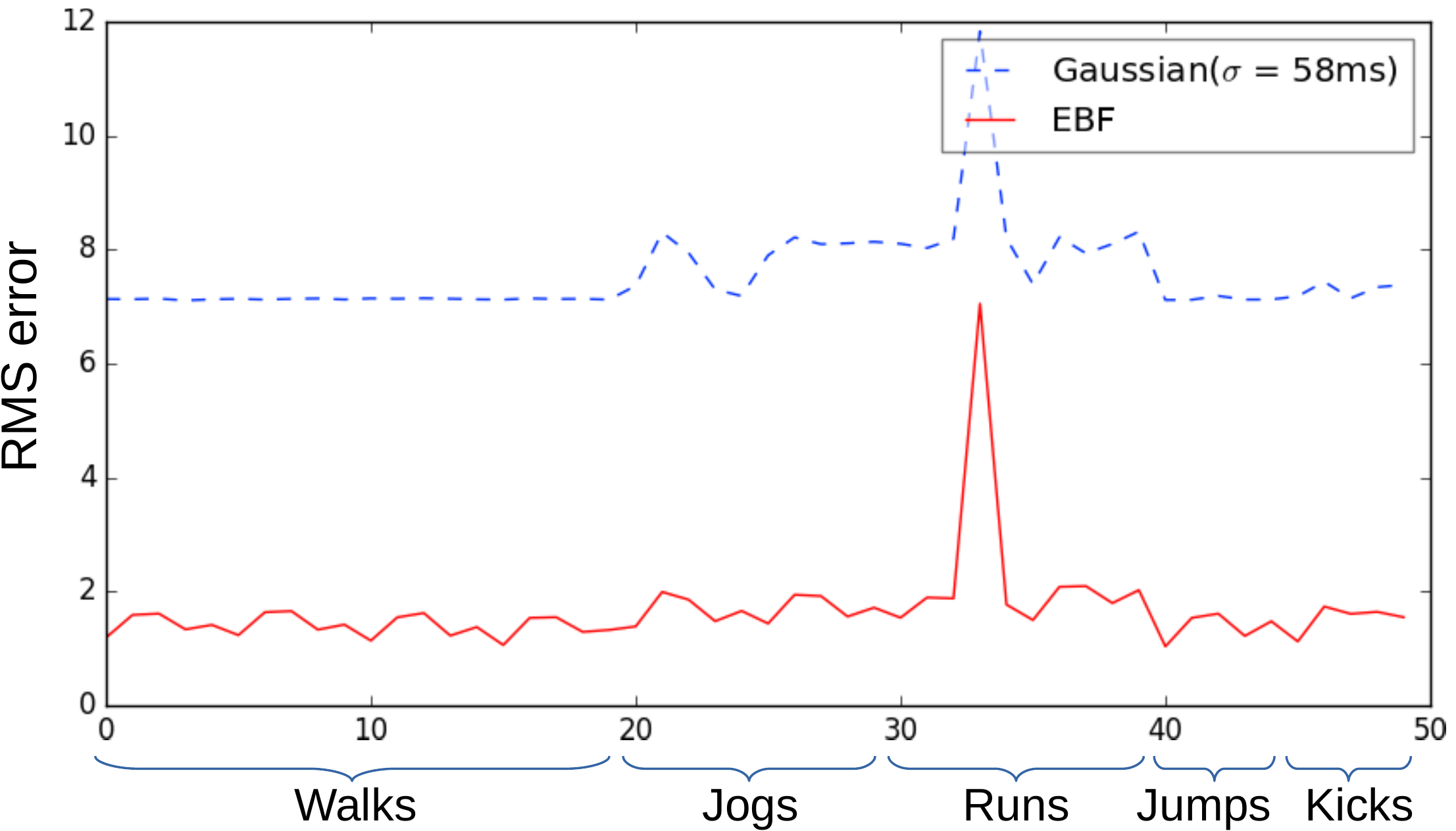}
    \caption{Constant bias plus 0.5 Gaussian}
    \label{fig:expt3_const}
  \end{subfigure}~%
  \begin{subfigure}[b]{0.32\linewidth}
    \includegraphics[width=\columnwidth]{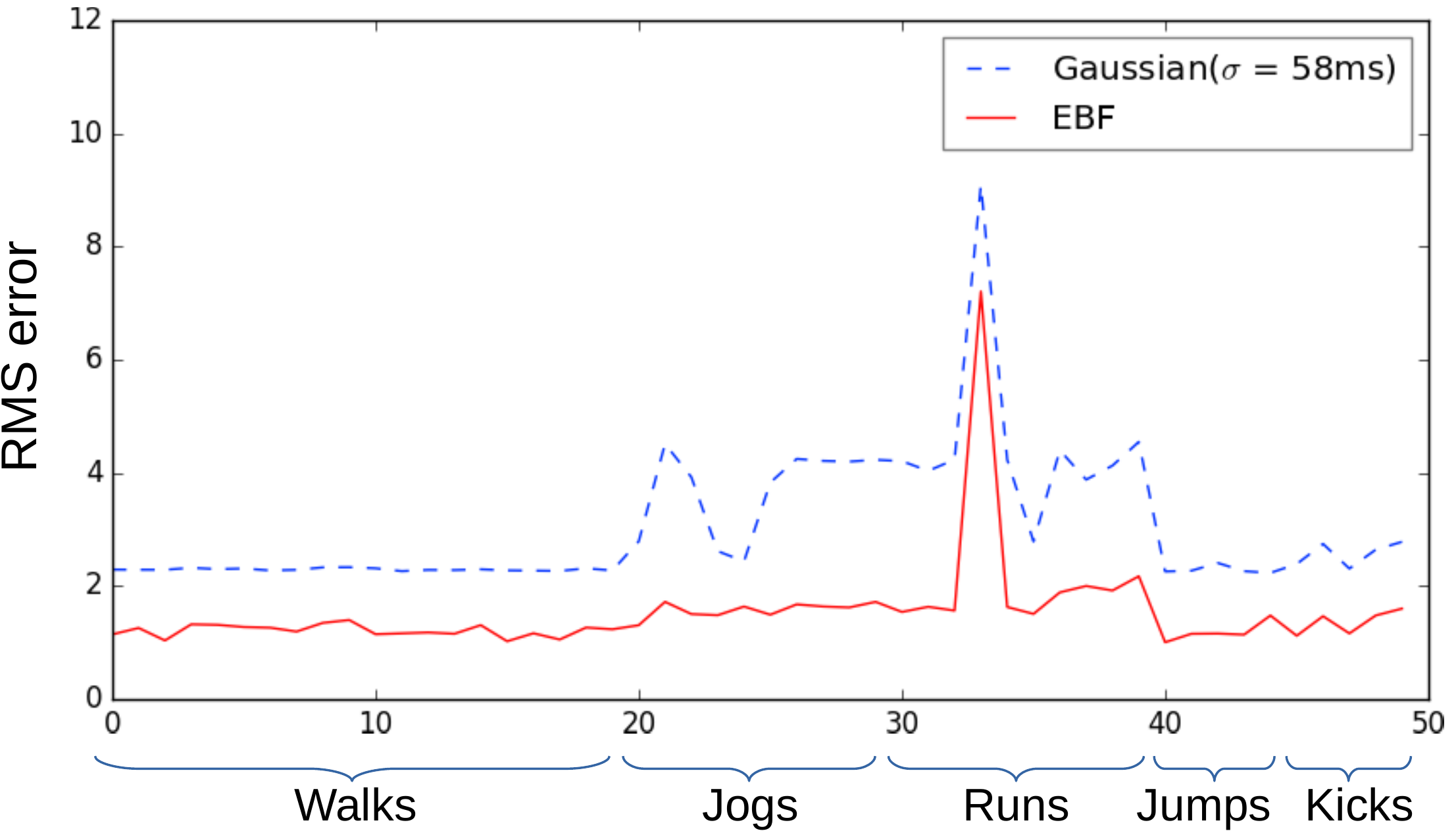}
    \caption{Sine wave plus 0.5 Gaussian}
    \label{fig:expt3_sine}
  \end{subfigure}
  \caption{Denoising performance of our EBF model on different noise distributions.}
  \label{fig:expt3}
\end{figure*}

\paragraph{Comparison of noise amplitudes.} We train multiple EBF models with different noise amplitudes. Three EBF models are trained on datasets that have $0.1$, $0.5$ and $0.9$ noise respectively. We also train a single EBF with all the noisy motions used to train the previously mentioned variants. Then we test the performance of these models on each of the different noisy datasets, and plot the RMS error, averaged over all frames, for each motion from all the holdouts. It can be seen in Figure~\ref{fig:expt2} (left) that for each noise amplitude, all EBF models perform almost similarly. Therefore, unless specified otherwise, we have used an EBF trained with $0.5$ noise in all our other experiments.

We also compare the denoising performance of our EBF models, trained and tested on $0.1$ ($20$dB SNR) and $0.9$ ($1$dB SNR) noise respectively, with a baseline Gaussian filter. A Gaussian with higher standard deviation is used as the baseline for higher noise since it was found to be more accurate. Similarly, we also compare the performance of EBF with a Gaussian baseline on different amplitudes of spatial noise. As can be seen in Figure~\ref{fig:expt2} (middle and right) EBF always outperforms the baseline.

\paragraph{Performance on different noise types.} We test the performance of our EBF model (with bias outputs) on uniform noise; on $0.5$ noise plus a constant bias; and on $0.5$ noise plus a sinusoidally varying bias. The EBF is trained and tested on data containing each kind of noise in turn, and compared to a Gaussian baseline. Figure~\ref{fig:expt3} shows that the EBF is able to successfully learn from data containing different types of noise, and is able to subsequently denoise it. Note that for bias-free distributions, the EBF-with-bias model learnt an almost exactly zero bias term, as expected.

\begin{figure}[b!]
\centering
  \includegraphics[width=\columnwidth]{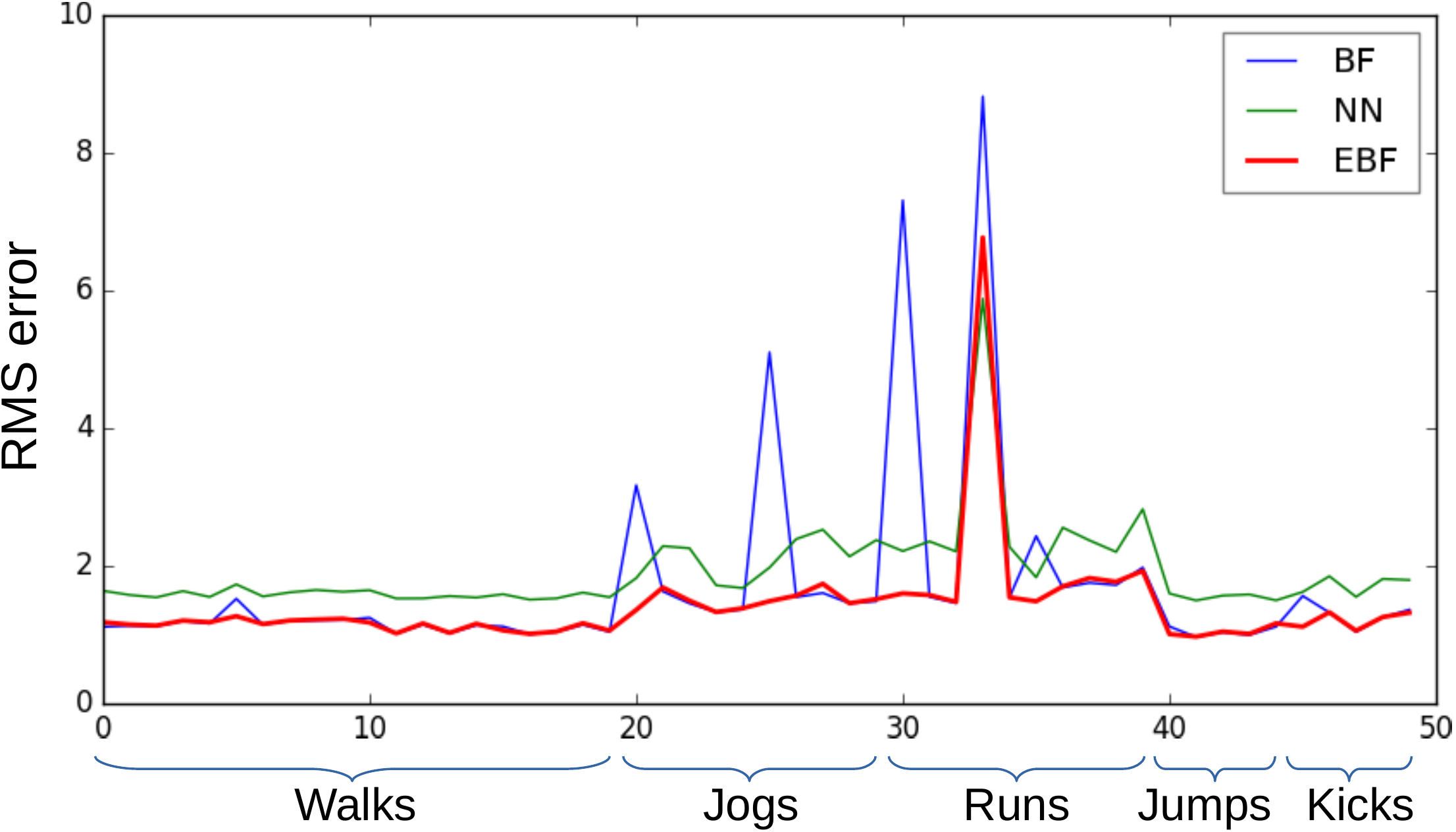}
  \caption{Denoising performance of networks with different architectures. The E and B layers of the EBF network enable it to perform better than variants that omit them.}
  \label{fig:expt4}
\end{figure}

\paragraph{Ablation studies with architecture variants.} We compare $3$ different network architectures. In the first architecture (denoted BF) we drop the E (Encoder) layer and train a network with only the B and F layers. In the second architecture (denoted NN) we train a feedforward network with $6$ fully connected dense layers, with $\tanh$ activation in all layers except the final one. The input is a single vector of size $3199 = 129 \times 31$ and the output is a vector of values for $126$ joint channels. The third network is our EBF model. The variation in average RMS error for all test motions (Figure~\ref{fig:expt4}) shows that EBF (avg error {\bf 1.46}) performs better at denoising than both the BF ($1.70$) and NN ($1.92$) networks. This indicates that the encoder module and the recurrent architecture are both important.

\paragraph{Extrapolation to other actions.} To test whether our framework trained on one type of motion can extrapolate to cleaning a different type, we held out each of the $5$ action categories (walk, jog, run, jump, kick) in turn for testing, and trained on the remaining $4$ categories. In this test, the average RMS error over the $50$ motions selected in the original holdouts is $1.76$. This compares favorably with Gaussian smoothing ($2.25$) but is expectedly not as good as when the EBF is trained on all action types ({\bf 1.42}).

\subsection{Gap Filling Results}
\begin{figure}[t!]
\centering
  \includegraphics[width=\columnwidth]{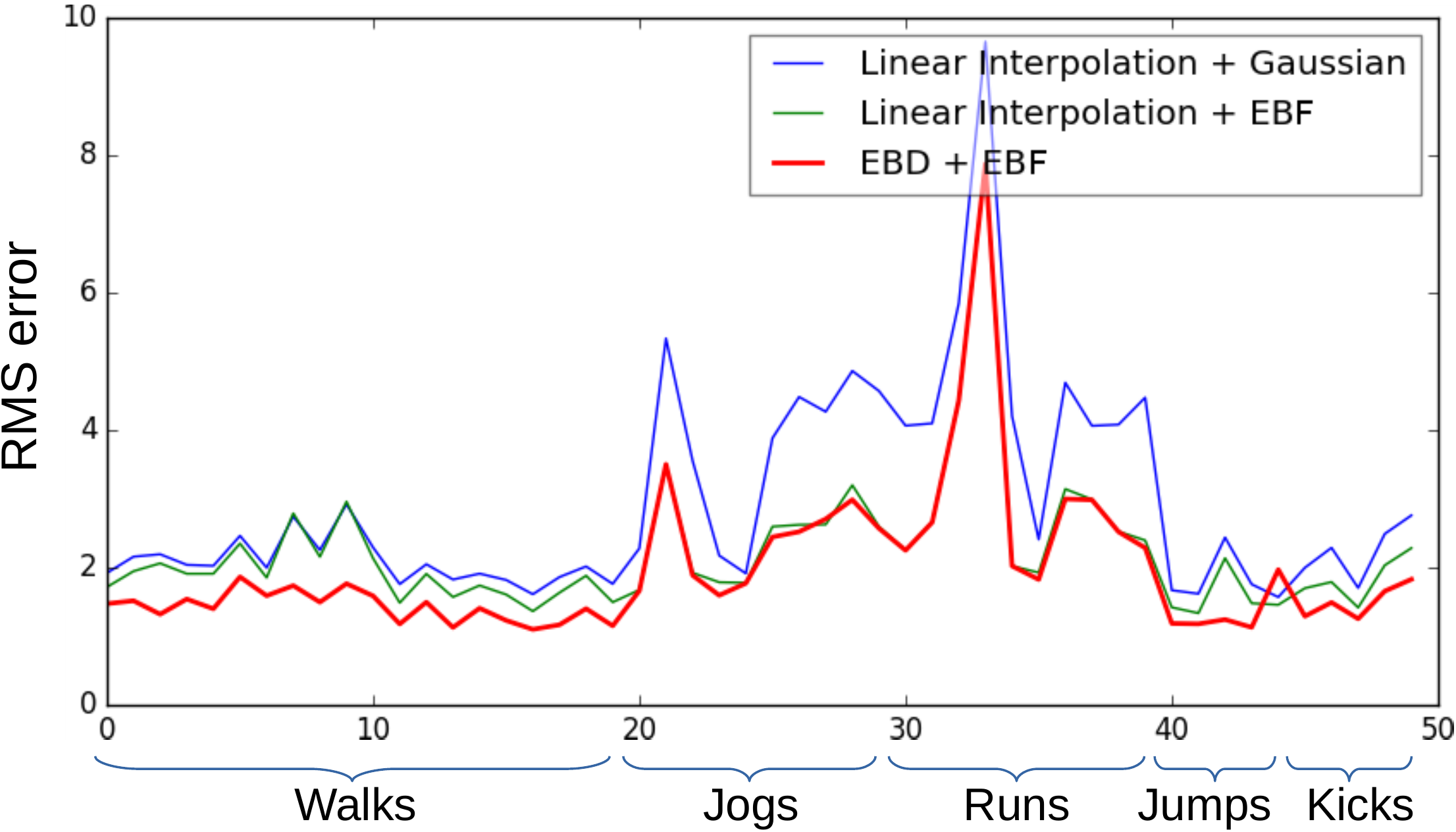}
  \caption{A comparison of three different combinations of gap filling and motion denoising methods for all the motions in our test holdouts. For each motion, the RMS error has been averaged over all frames. EBF + EBD performs consistently better than other methods.}
  \label{fig:expt5_2}
\end{figure}

We look at the performance of our EBF + EBD framework in the presence of missing samples. We train the networks on noisy data with missing samples. Runs of missing samples or gaps are generated in the data preparation step as explained above. We compare the performance of three methods for cleaning motion capture data with gaps and noise. In the first method, we use linear interpolation (lerp) to close the gaps, and then use a Gaussian filter to denoise the complete motion. The second method again uses lerp to close the gaps but uses an EBF to denoise it. The third method uses the trained EBD network to predict the missing samples and then uses the EBF to denoise it. Figure~\ref{fig:expt5_2} presents the RMS error averaged over all frames of each motion from the $5$ holdouts. The combination of using EBD for gap filling, followed by EBF for denoising, performs the best for nearly all test motions.

\begin{figure}[h!]
 \vspace{4mm}
\centering
  \begin{subfigure}[b]{0.98\columnwidth}
    \includegraphics[width=\columnwidth]{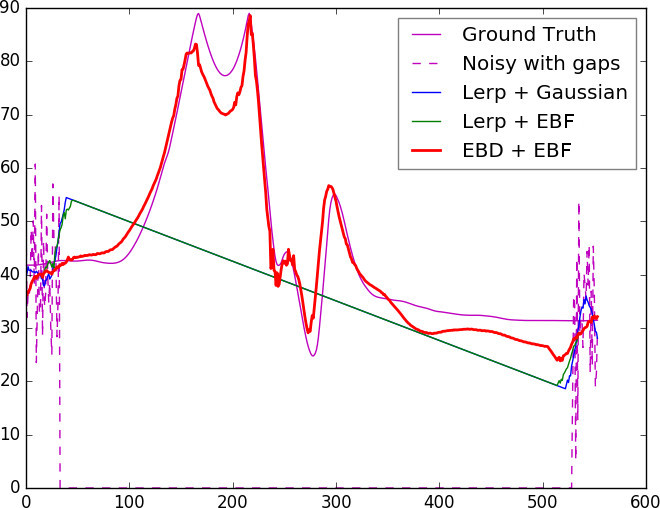}
  \end{subfigure}~\\ \vspace{2mm}
  \begin{subfigure}[b]{0.48\columnwidth}
    \includegraphics[width=\columnwidth]{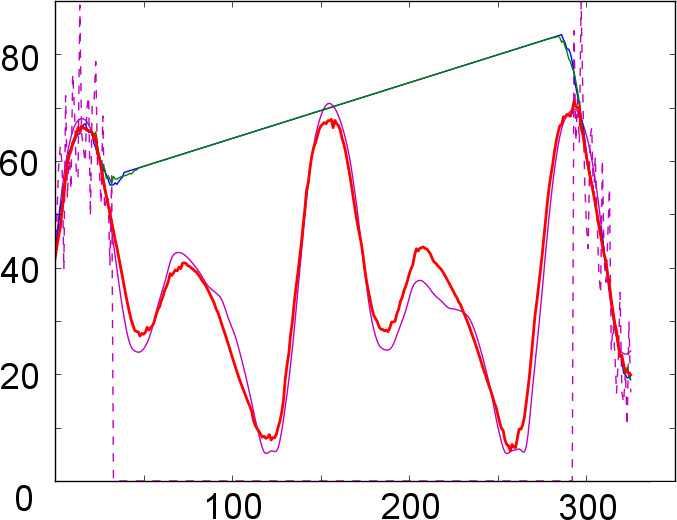}
  \end{subfigure}~%
  \begin{subfigure}[b]{0.48\columnwidth}
    \includegraphics[width=\columnwidth]{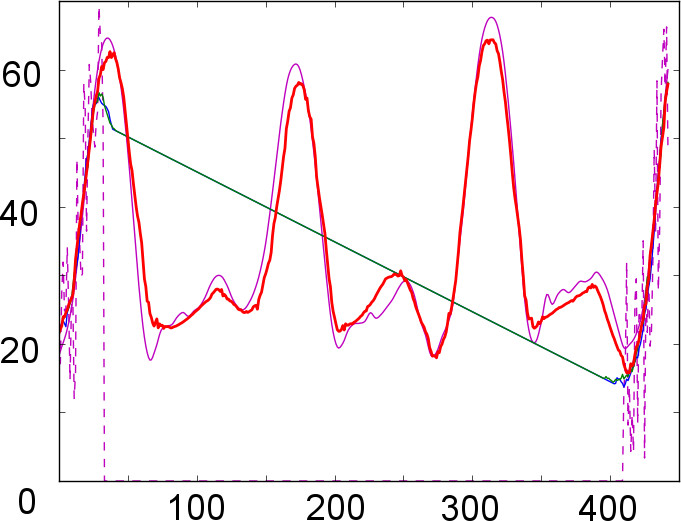}
  \end{subfigure}~%
  \caption{Reconstruction of missing samples in channels corresponding to rotation about X axis in the left knee joint (top and bottom right), and the right knee joint (bottom left). Very long gaps of 2-4 seconds were artificially introduced in these channels. Among all the compared methods, EBD+EBF is the only method that can close long gaps and reconstruct motions that closely resemble the ground truth. The blue line (linear interpolation plus Gaussian) is almost coincident with the green one.}
  \label{fig:expt5_4}
\end{figure}

Longer gaps, though occurring less frequently, are much harder to fill convincingly than smaller gaps. We specifically tested our gap filling method on data in which we introduced long gaps of up to $5$ seconds (i.e. 600 frames). The combination of EBD and EBF was able to successfully reconstruct the missing data in all our tests, whereas both the other methods mostly failed to do so. Examples of these long gap reconstructions can be seen in Figure~\ref{fig:expt5_4}. The remarkably accurate reconstructions can be attributed to learning correlations of the missing joint channel with all other joint channels in the skeleton, as well as temporal coherence.

\subsection{Timing Results}

We used TensorFlow to implement our networks, and trained on a variety of nVidia GPUs. The EBF model takes about $5$ hours and the EBD model $4$ hours to train for a single holdout on one GPU. Each frame of a motion clip can be denoised in real-time (\mbox{$<1$ms} per frame), allowing live processing of streamed motion capture data (with a short delay for acquiring the lookahead frames).
\section{Conclusion}
\label{sec:conclusion}

We present a deep recurrent framework for cleaning motion capture data with noisy and missing samples. The approach is not specific to a particular kind of noise or action, and requires no manual tuning. Our Encoder-Bidirectional-Filter (EBF) architecture predicts an adaptive low-pass filter for each joint channel in every frame. The network implicitly builds a model for temporal coherence and correlation between joint channels and can thus predict filters that adapt to account for large amounts of different noise types. To reconstruct missing samples in noisy motions, we use an Encoder-Bidirectional-Decoder (EBD) network. The result is then piped through the EBF network for denoising. We present an extensive set of experiments that measure and validate the performance of our methods. In particular, we illustrate the framework's ability to work with noise from different distributions, including additive and high amplitude noise, and in the presence of very large gaps.

Since the training is supervised, the network needs to see sufficient data of a new motion type before it can learn to clean it reliably. Although experiments suggest the network can extrapolate to motion types it has never seen \mbox{(Section \ref{sec:results}),} the quality of the cleaned data will certainly improve if the network has seen the motion during training. We were able to train a single network on all the different motions together. However, in order to make a generic argument about how much diversity a single network can handle, more testing with a wide variety of motions is needed. Similarly, we cannot make strong claims about the ability of a trained network to generalize to unseen noise distributions. Our method does not explicitly consider the lengths of bones that connect the joints in the skeleton, therefore it should be able to clean motion captured from a different performer (than the one used to generate the training data), as long as the topology of the recovered skeleton stays the same. We would also like to try and clean motion data obtained from a variety of motion capture systems like markerless systems based on RGBD cameras or inertial sensors, and extend the approach to general time series data.

We hope our method can serve as a default push-button solution for cleaning motion data, thereby saving valuable time in the motion capture data processing pipeline.

\bibliographystyle{ieee}
\bibliography{ebf}

\begin{thebibliography}{10}\itemsep=-1pt

\bibitem{akhter2012}
I.~Akhter, T.~Simon, S.~Khan, I.~Matthews, and Y.~Sheikh.
\newblock Bilinear spatiotemporal basis models.
\newblock {\em Trans. Graphics}, 31(2):17:1--17:12, 2012.

\bibitem{aristidou2013}
A.~Aristidou and J.~Lasenby.
\newblock Real-time marker prediction and {CoR} estimation in optical motion
  capture.
\newblock {\em The Visual Computer}, 29(1):7--26, 2013.

\bibitem{maya2017}
Autodesk.
\newblock Maya.
\newblock \url{http://www.autodesk.com/products/maya/overview}, 2017.

\bibitem{bako2017}
S.~Bako, T.~Vogels, B.~Mcwilliams, M.~Meyer, J.~Nov\'{a}K, A.~Harvill, P.~Sen,
  T.~Derose, and F.~Rousselle.
\newblock Kernel-predicting convolutional networks for denoising {M}onte
  {C}arlo renderings.
\newblock {\em Trans. Graph.}, 36(4), 2017.

\bibitem{barker2016}
J.~Barker, R.~Marxer, E.~Vincent, and S.~Watanabe.
\newblock The third {CHiME} speech separation and recognition challenge:
  Analysis and outcomes.
\newblock {\em Computer Speech \& Language}, 2016.

\bibitem{blender2017}
BlenderFoundation.
\newblock Blender.
\newblock \url{https://www.blender.org/}, 2017.

\bibitem{burke2016}
M.~Burke and J.~Lasenby.
\newblock Estimating missing marker positions using low dimensional {Kalman}
  smoothing.
\newblock {\em J. Biomechanics}, 49(9):1854--1858, 2016.

\bibitem{cmu2017}
CMU.
\newblock {CMU} {G}raphics {L}ab {M}otion {C}apture {D}atabase.
\newblock \url{http://mocap.cs.cmu.edu/}, 2017.

\bibitem{dorfmuller2003}
K.~Dorfm{\"u}ller-Ulhaas.
\newblock Robust optical user motion tracking using a {Kalman} filter.
\newblock In {\em VRST}, 2003.

\bibitem{du2015}
Y.~Du, W.~Wang, and L.~Wang.
\newblock Hierarchical recurrent neural network for skeleton based action
  recognition.
\newblock In {\em CVPR}, pages 1110--1118, 2015.

\bibitem{mousa2015}
A.~El-Desoky~Mousa, E.~Marchi, and B.~Schuller.
\newblock The {ICSTM+} {TUM+} {UP} approach to the 3rd {CHiME} challenge:
  Single-channel {LSTM} speech enhancement with multi-channel correlation
  shaping dereverberation and {LSTM} language models.
\newblock {\em arXiv preprint arXiv:1510.00268}, 2015.

\bibitem{fragkiadaki2015}
K.~Fragkiadaki, S.~Levine, P.~Felsen, and J.~Malik.
\newblock Recurrent network models for human dynamics.
\newblock In {\em ICCV}, pages 4346--4354, 2015.

\bibitem{graves2005}
A.~Graves and J.~Schmidhuber.
\newblock Framewise phoneme classification with bidirectional {LSTM} and other
  neural network architectures.
\newblock {\em Neural Networks}, 18(5):602--610, 2005.

\bibitem{herda2000}
L.~Herda, P.~Fua, R.~Plankers, R.~Boulic, and D.~Thalmann.
\newblock Skeleton-based motion capture for robust reconstruction of human
  motion.
\newblock In {\em Computer Animation}, pages 77--83, 2000.

\bibitem{holden2017}
D.~Holden, T.~Komura, and J.~Saito.
\newblock Phase-functioned neural networks for character control.
\newblock {\em Trans. Graph.}, 36(4), 2017.

\bibitem{holden2016}
D.~Holden, J.~Saito, and T.~Komura.
\newblock A deep learning framework for character motion synthesis and editing.
\newblock {\em Trans. Graphics}, 35(4), 2016.

\bibitem{holden2015}
D.~Holden, J.~Saito, T.~Komura, and T.~Joyce.
\newblock Learning motion manifolds with convolutional autoencoders.
\newblock In {\em SIGGRAPH Asia Technical Briefs}, page~18, 2015.

\bibitem{hornung2005}
A.~Hornung, S.~Sar-Dessai, and L.~Kobbelt.
\newblock Self-calibrating optical motion tracking for articulated bodies.
\newblock In {\em Virtual Reality}, pages 75--82, 2005.

\bibitem{li2010}
L.~Li, J.~McCann, N.~Pollard, and C.~Faloutsos.
\newblock {BoLeRO}: a principled technique for including bone length
  constraints in motion capture occlusion filling.
\newblock In {\em SCA}, pages 179--188, 2010.

\bibitem{liu2006}
G.~Liu and L.~McMillan.
\newblock Estimation of missing markers in human motion capture.
\newblock {\em The Visual Computer}, 22(9):721--728, 2006.

\bibitem{lou2010}
H.~Lou and J.~Chai.
\newblock Example-based human motion denoising.
\newblock {\em TVCG}, 16(5):870--879, 2010.

\bibitem{organic2017}
OrganicMotion.
\newblock \url{http://www.organicmotion.com/}, 2017.

\bibitem{park2006}
S.~I. Park and J.~K. Hodgins.
\newblock Capturing and animating skin deformation in human motion.
\newblock {\em Trans. Graphics}, 25(3):881--889, 2006.

\bibitem{taylor2007}
G.~W. Taylor, G.~E. Hinton, and S.~T. Roweis.
\newblock Modeling human motion using binary latent variables.
\newblock {\em NIPS}, 19:1345, 2007.

\bibitem{vicon2017}
Vicon.
\newblock \url{https://www.vicon.com/}, 2017.

\bibitem{weninger2015}
F.~Weninger, H.~Erdogan, S.~Watanabe, E.~Vincent, J.~Le~Roux, J.~R. Hershey,
  and B.~Schuller.
\newblock Speech enhancement with {LSTM} recurrent neural networks and its
  application to noise-robust {ASR}.
\newblock In {\em Intl. Conf. Latent Variable Analysis and Signal Separation},
  pages 91--99, 2015.

\bibitem{xiao2015}
J.~Xiao, Y.~Feng, M.~Ji, X.~Yang, J.~J. Zhang, and Y.~Zhuang.
\newblock Sparse motion bases selection for human motion denoising.
\newblock {\em Signal Processing}, 110:108--122, 2015.

\bibitem{zordan2003}
V.~B. Zordan and N.~C. Van Der~Horst.
\newblock Mapping optical motion capture data to skeletal motion using a
  physical model.
\newblock In {\em SCA}, pages 245--250, 2003.

\end{thebibliography}

\end{document}